\documentclass[conference,comsoc]{IEEEtran}
\IEEEoverridecommandlockouts

\usepackage{cite}
\usepackage{amsmath,amssymb,amsfonts}
\usepackage{algorithm,algorithmic}
\usepackage{subcaption}
\usepackage{graphicx}
\usepackage{textcomp}
\usepackage{xcolor}
\def\BibTeX{{\rm B\kern-.05em{\sc i\kern-.025em b}\kern-.08em
    T\kern-.1667em\lower.7ex\hbox{E}\kern-.125emX}}

\begin{document}
\bstctlcite{IEEEexample:BSTcontrol}

\title{Switch-DFT: Adaptive Waveform and MIMO Switching for Energy-Efficient Base Stations}

\author{\IEEEauthorblockN{Jaebum Park$^{*}$,~Chan-Byoung Chae$^{\dagger}$, and Robert W. Heath Jr.$^{*}$}
\IEEEauthorblockA{$^{*}$Department of Electrical and Computer Engineering, UC San Diego, La Jolla, CA, 92093 USA \\
$^{\dagger}$School of Integrated Technology, Yonsei University, Seoul, 03722 South Korea\\E-mail: \{jp261, rwheathjr\}@ucsd.edu; cbchae@yonsei.ac.kr }}
\maketitle

\newcommand{\ue}{u \color{black}}
\newcommand{\ssb}{i \color{black}}
\newcommand{\pathl}{\ell \color{black}}
\newcommand{\subc}{k \color{black}}
\newcommand{\timet}{t \color{black}}
\newcommand{\bandK}{K \color{black}}
\newcommand{\timeT}{T \color{black}}
\newcommand{\ues}{U \color{black}}

\newcommand{\analog}[1]{#1^{\text{RF}}}
\newcommand{\Nxa}{\analog{N}_{\text{X}}}
\newcommand{\Nya}{\analog{N}_{\text{Y}}}
\newcommand{\Nxd}{N_{\text{Xd}}}
\newcommand{\Nyd}{N_{\text{Yd}}}
\newcommand{\Ntd}{N_{\text{Td}}}

\newcommand{\Nx}{N_{\text{X}}}
\newcommand{\Ny}{N_{\text{Y}}}
\newcommand{\Nt}{N_{\text{T}}}
\newcommand{\Nta}{\analog{N}_{\text{T}}}
\newcommand{\Nr}{N_{\text{R}}}
\newcommand{\Nra}{\analog{N}_{\text{R}}}
\newcommand{\nt}{n_{\text{t}}}
\newcommand{\nr}{n_{\text{r}}}
\newcommand{\Np}{N_{\text{P}}}
\newcommand{\Oh}{O_{\text{H}}}
\newcommand{\Ov}{O_{\text{V}}}

\newcommand{\Real}{\mathbb{R}}
\newcommand{\Imag}{\mathbb{I}}
\newcommand{\norm}[1]{\left\lVert#1\right\rVert}
\newcommand{\setof}[1]{\{#1\}_{n=0}^{N-1}}
\newcommand{\ji}{\mathsf{j}}
\newcommand{\pathgain}{\alpha_{\pathl}}
\newcommand{\phaseshift}{\beta_{\pathl}}
\newcommand{\azimuth}[1]{\theta_{\pathl}^{#1}}
\newcommand{\elevation}[1]{\phi_{\pathl}^{#1}}
\newcommand{\vander}[2]{\mathbf{a}_{N_{#2}}(#1)}
\newcommand{\vect}[1]{\mathbf{#1}}
\newcommand{\complex}{\mathbb{C}}
\newcommand{\steering}[1]{\mathbf{A}(#1)}
\newcommand{\conjsteering}[1]{\mathbf{A}^*(#1)}
\newcommand{\phaseres}{b_{\text{phase}}}
\newcommand{\w}{\mathbf{w}}
\newcommand{\W}{\mathbf{W}}
\newcommand{\Wutk}{\mathbf{W}_{\ue, t, k}}
\newcommand{\F}{\mathbf{F}}
\newcommand{\Futk}{\mathbf{F}_{\ue, t, k}}
\newcommand{\f}{\mathbf{f}}
\newcommand{\Wa}{\analog{\mathbf{W}}_{\ue, t}}
\newcommand{\Fa}{\analog{\mathbf{F}}_{c, t}}
\newcommand{\Fap}{\analog{\mathbf{F}}_{c', t}}
\newcommand{\fa}{\analog{\mathbf{f}}_{t}}
\newcommand{\Fistar}{\F_{b_u, \hat{i}_{b_u, u}}}
\newcommand{\Fssb}{\mathbf{F}^{\text{SSB}}}
\newcommand{\Fcsi}{\mathbf{F}^{\text{CSI-RS}}}
\newcommand{\Bssb}{\mathbf{B}^{\text{SSB}}}
\newcommand{\Bcsi}{\mathbf{B}^{\text{CSI-RS}}}
\newcommand{\Bdft}{\mathbf{B}^{\text{DFT}}}
\newcommand{\Bactive}{\mathbf{B}^{\text{sub}}}
\newcommand{\Str}{\vect{s}^{\text{tr}}_{c, t, k}}
\newcommand{\Lmax}{L_{\text{{max}}}}
\newcommand{\Kssb}{K^{\text{SSB}}}
\newcommand{\Tssb}{T^{\text{SSB}}_{i}}
\newcommand{\yssb}{y^{\text{SSB}}_{\ue, i}}
\newcommand{\ycsirs}{\vect{y}^{\text{CSI-RS}_i}}
\newcommand{\Kcsi}{K^{\text{CSI-RS}_i}}
\newcommand{\Tcsi}{T^{\text{CSI-RS}_{i}}}
\newcommand{\Lcsi}{L_{\text{{CSI}}}}
\newcommand{\bwp}{\text{S}_{\text{B}}}
\newcommand{\Pcsi}{P_{\text{CSI}}}
\newcommand{\nRB}{N_{\text{RB}}}
\newcommand{\Ncsi}{N_{\text{CSI}}}
\newcommand{\Bg}{B_{\text{g}}}
\newcommand{\ihat}{\widehat{\vect{i}}_{c, u}}
\newcommand{\ihati}{\widehat{\vect{i}}_{c, i}}
 \newcommand{\Obsc}{\vect{O}^{\text{BSC}}}
 \newcommand{\Nxo}{N_{\text{X,O}}}
  \newcommand{\Nyo}{N_{\text{Y,O}}}
  \newcommand{\Fssbpred}{\widehat{\mathbf{F}}^{\text{SSB}}}
\newcommand{\Fcsipred}{\widehat{\mathbf{F}}^{\text{CSI-RS}}}
\newcommand{\Achan}{\vect{H}}
\newcommand{\Dchan}{\vect{H}^{\text{BB}}}
\newcommand{\chanNT}{\mathbf{H}}
\newcommand{\Hfftu}[1]{\boldsymbol{\mathit{H}^{(\ue)}_{#1}}}
\newcommand{\Hall}{\boldsymbol{\mathit{H}}}

\newcommand{\bigchan}{\mathbf{\mathcal{H}}}
\newcommand{\Uactive}{U_{\text{active}}}
\newcommand{\usel}[1]{#1_{(\ue)}}
\newcommand{\algo}{\text{NBL}}
\newcommand{\tku}[1]{{#1_{\ue, t, k}}}

\newcommand{\RSRPiu}{\text{RSRP}_{i, u}}
\newcommand{\RSRPciu}{\text{RSRP}_{c, i, u}}
\newcommand{\SNR}{\text{SNR}}
\newcommand{\SINR}{\text{SINR}}
\newcommand{\EVM}{\text{EVM}}
\newcommand{\PA}{\text{PA}}
\newcommand{\mix}{\text{mix}}
\newcommand{\LO}{\text{LO}}
\newcommand{\filt}{\text{filt}}
\newcommand{\dac}{\text{dac}}
\newcommand{\idle}{\text{idle}}
\newcommand{\sat}{\text{sat}}
\newcommand{\TX}{\text{TX}}
\newcommand{\RX}{\text{RX}}
\newcommand{\RU}{\text{RU}}

\newcommand{\figref}[1]{Fig.~\ref{#1}}
\newcommand{\tabref}[1]{Table~\ref{#1}}
\newcommand{\secref}[1]{Section~\ref{#1}}
\newcommand{\algref}[1]{Algorithm~\ref{#1}}
\def\bydef{:=}
\def\bba{{\mathbb{a}}}
\def\bbb{{\mathbb{b}}}
\def\bbc{{\mathbb{c}}}
\def\bbd{{\mathbb{d}}}
\def\bbee{{\mathbb{e}}}
\def\bbff{{\mathbb{f}}}
\def\bbg{{\mathbb{g}}}
\def\bbh{{\mathbb{h}}}
\def\bbi{{\mathbb{i}}}
\def\bbj{{\mathbb{j}}}
\def\bbk{{\mathbb{k}}}
\def\bbl{{\mathbb{l}}}
\def\bbm{{\mathbb{m}}}
\def\bbn{{\mathbb{n}}}
\def\bbo{{\mathbb{o}}}
\def\bbp{{\mathbb{p}}}
\def\bbq{{\mathbb{q}}}
\def\bbr{{\mathbb{r}}}
\def\bbs{{\mathbb{s}}}
\def\bbt{{\mathbb{t}}}
\def\bbu{{\mathbb{u}}}
\def\bbv{{\mathbb{v}}}
\def\bbw{{\mathbb{w}}}
\def\bbx{{\mathbb{x}}}
\def\bby{{\mathbb{y}}}
\def\bbz{{\mathbb{z}}}
\def\bb0{{\mathbb{0}}}

\def\bydef{:=}
\def\ba{{\mathbf{a}}}
\def\bb{{\mathbf{b}}}
\def\bc{{\mathbf{c}}}
\def\bd{{\mathbf{d}}}
\def\bee{{\mathbf{e}}}
\def\bff{{\mathbf{f}}}
\def\bg{{\mathbf{g}}}
\def\bh{{\mathbf{h}}}
\def\bi{{\mathbf{i}}}
\def\bj{{\mathbf{j}}}
\def\bk{{\mathbf{k}}}
\def\bl{{\mathbf{l}}}
\def\bm{{\mathbf{m}}}
\def\bn{{\mathbf{n}}}
\def\bo{{\mathbf{o}}}
\def\bp{{\mathbf{p}}}
\def\bq{{\mathbf{q}}}
\def\br{{\mathbf{r}}}
\def\bs{{\mathbf{s}}}
\def\bt{{\mathbf{t}}}
\def\bu{{\mathbf{u}}}
\def\bv{{\mathbf{v}}}
\def\bw{{\mathbf{w}}}
\def\bx{{\mathbf{x}}}
\def\by{{\mathbf{y}}}
\def\bz{{\mathbf{z}}}
\def\b0{{\mathbf{0}}}

\def\bA{{\mathbf{A}}}
\def\bB{{\mathbf{B}}}
\def\bC{{\mathbf{C}}}
\def\bD{{\mathbf{D}}}
\def\bE{{\mathbf{E}}}
\def\bF{{\mathbf{F}}}
\def\bG{{\mathbf{G}}}
\def\bH{{\mathbf{H}}}
\def\bI{{\mathbf{I}}}
\def\bJ{{\mathbf{J}}}
\def\bK{{\mathbf{K}}}
\def\bL{{\mathbf{L}}}
\def\bM{{\mathbf{M}}}
\def\bN{{\mathbf{N}}}
\def\bO{{\mathbf{O}}}
\def\bP{{\mathbf{P}}}
\def\bQ{{\mathbf{Q}}}
\def\bR{{\mathbf{R}}}
\def\bS{{\mathbf{S}}}
\def\bT{{\mathbf{T}}}
\def\bU{{\mathbf{U}}}
\def\bV{{\mathbf{V}}}
\def\bW{{\mathbf{W}}}
\def\bX{{\mathbf{X}}}
\def\bY{{\mathbf{Y}}}
\def\bZ{{\mathbf{Z}}}


\def\bbA{{\mathbb{A}}}
\def\bbB{{\mathbb{B}}}
\def\bbC{{\mathbb{C}}}
\def\bbD{{\mathbb{D}}}
\def\bbE{{\mathbb{E}}}
\def\bbF{{\mathbb{F}}}
\def\bbG{{\mathbb{G}}}
\def\bbH{{\mathbb{H}}}
\def\bbI{{\mathbb{I}}}
\def\bbJ{{\mathbb{J}}}
\def\bbK{{\mathbb{K}}}
\def\bbL{{\mathbb{L}}}
\def\bbM{{\mathbb{M}}}
\def\bbN{{\mathbb{N}}}
\def\bbO{{\mathbb{O}}}
\def\bbP{{\mathbb{P}}}
\def\bbQ{{\mathbb{Q}}}
\def\bbR{{\mathbb{R}}}
\def\bbS{{\mathbb{S}}}
\def\bbT{{\mathbb{T}}}
\def\bbU{{\mathbb{U}}}
\def\bbV{{\mathbb{V}}}
\def\bbW{{\mathbb{W}}}
\def\bbX{{\mathbb{X}}}
\def\bbY{{\mathbb{Y}}}
\def\bbZ{{\mathbb{Z}}}


\def\cA{\mathcal{A}}
\def\cB{\mathcal{B}}
\def\cC{\mathcal{C}}
\def\cD{\mathcal{D}}
\def\cE{\mathcal{E}}
\def\cF{\mathcal{F}}
\def\cG{\mathcal{G}}
\def\cH{\mathcal{H}}
\def\cI{\mathcal{I}}
\def\cJ{\mathcal{J}}
\def\cK{\mathcal{K}}
\def\cL{\mathcal{L}}
\def\cM{\mathcal{M}}
\def\cN{\mathcal{N}}
\def\cO{\mathcal{O}}
\def\cP{\mathcal{P}}
\def\cQ{\mathcal{Q}}
\def\cR{\mathcal{R}}
\def\cS{\mathcal{S}}
\def\cT{\mathcal{T}}
\def\cU{\mathcal{U}}
\def\cV{\mathcal{V}}
\def\cW{\mathcal{W}}
\def\cX{\mathcal{X}}
\def\cY{\mathcal{Y}}
\def\cZ{\mathcal{Z}}

\def\sfA{\mathsf{A}}
\def\sfB{\mathsf{B}}
\def\sfC{\mathsf{C}}
\def\sfD{\mathsf{D}}
\def\sfE{\mathsf{E}}
\def\sfF{\mathsf{F}}
\def\sfG{\mathsf{G}}
\def\sfH{\mathsf{H}}
\def\sfI{\mathsf{I}}
\def\sfJ{\mathsf{J}}
\def\sfK{\mathsf{K}}
\def\sfL{\mathsf{L}}
\def\sfM{\mathsf{M}}
\def\sfN{\mathsf{N}}
\def\sfO{\mathsf{O}}
\def\sfP{\mathsf{P}}
\def\sfQ{\mathsf{Q}}
\def\sfR{\mathsf{R}}
\def\sfS{\mathsf{S}}
\def\sfT{\mathsf{T}}
\def\sfU{\mathsf{U}}
\def\sfV{\mathsf{V}}
\def\sfW{\mathsf{W}}
\def\sfX{\mathsf{X}}
\def\sfY{\mathsf{Y}}
\def\sfZ{\mathsf{Z}}

\def\bydef{:=}
\def\sfa{{\mathsf{a}}}
\def\sfb{{\mathsf{b}}}
\def\sfc{{\mathsf{c}}}
\def\sfd{{\mathsf{d}}}
\def\sfee{{\mathsf{e}}}
\def\sfff{{\mathsf{f}}}
\def\sfg{{\mathsf{g}}}
\def\sfh{{\mathsf{h}}}
\def\sfi{{\mathsf{i}}}
\def\sfj{{\mathsf{j}}}
\def\sfk{{\mathsf{k}}}
\def\sfl{{\mathsf{l}}}
\def\sfm{{\mathsf{m}}}
\def\sfn{{\mathsf{n}}}
\def\sfo{{\mathsf{o}}}
\def\sfp{{\mathsf{p}}}
\def\sfq{{\mathsf{q}}}
\def\sfr{{\mathsf{r}}}
\def\sfs{{\mathsf{s}}}
\def\sft{{\mathsf{t}}}
\def\sfu{{\mathsf{u}}}
\def\sfv{{\mathsf{v}}}
\def\sfw{{\mathsf{w}}}
\def\sfx{{\mathsf{x}}}
\def\sfy{{\mathsf{y}}}
\def\sfz{{\mathsf{z}}}
\def\sf0{{\mathsf{0}}}


\def\rmA{\mathrm{A}}
\def\rmB{\mathrm{B}}
\def\rmC{\mathrm{C}}
\def\rmD{\mathrm{D}}
\def\rmE{\mathrm{E}}
\def\rmF{\mathrm{F}}
\def\rmG{\mathrm{G}}
\def\rmH{\mathrm{H}}
\def\rmI{\mathrm{I}}
\def\rmJ{\mathrm{J}}
\def\rmK{\mathrm{K}}
\def\rmL{\mathrm{L}}
\def\rmM{\mathrm{M}}
\def\rmN{\mathrm{N}}
\def\rmO{\mathrm{O}}
\def\rmP{\mathrm{P}}
\def\rmQ{\mathrm{Q}}
\def\rmR{\mathrm{R}}
\def\rmS{\mathrm{S}}
\def\rmT{\mathrm{T}}
\def\rmU{\mathrm{U}}
\def\rmV{\mathrm{V}}
\def\rmW{\mathrm{W}}
\def\rmX{\mathrm{X}}
\def\rmY{\mathrm{Y}}
\def\rmZ{\mathrm{Z}}

\def\bydef{:=}
\def\rma{{\mathrm{a}}}
\def\rmb{{\mathrm{b}}}
\def\rmc{{\mathrm{c}}}
\def\rmd{{\mathrm{d}}}
\def\rmee{{\mathrm{e}}}
\def\rmff{{\mathrm{f}}}
\def\rmg{{\mathrm{g}}}
\def\rmh{{\mathrm{h}}}
\def\rmi{{\mathrm{i}}}
\def\rmj{{\mathrm{j}}}
\def\rmk{{\mathrm{k}}}
\def\rml{{\mathrm{l}}}
\def\rmm{{\mathrm{m}}}
\def\rmn{{\mathrm{n}}}
\def\rmo{{\mathrm{o}}}
\def\rmp{{\mathrm{p}}}
\def\rmq{{\mathrm{q}}}
\def\rmr{{\mathrm{r}}}
\def\rms{{\mathrm{s}}}
\def\rmt{{\mathrm{t}}}
\def\rmu{{\mathrm{u}}}
\def\rmv{{\mathrm{v}}}
\def\rmw{{\mathrm{w}}}
\def\rmx{{\mathrm{x}}}
\def\rmy{{\mathrm{y}}}
\def\rmz{{\mathrm{z}}}
\def\rm0{{\mathrm{0}}}

\definecolor{Gray}{gray}{0.85}
\definecolor{LightCyan}{rgb}{0.88,1,1}
\definecolor{ceruleanblue}{rgb}{0.16, 0.32, 0.75}
\definecolor{purple(x11)}{rgb}{0.63, 0.36, 0.94}
\definecolor{cadmiumgreen}{rgb}{0.0, 0.42, 0.24}
\definecolor{tortilla}{rgb}{0.597, 0.472, 0.3125}
\definecolor{amethyst}{rgb}{0.6, 0.4, 0.8}
\newcommand{\JP}[1]{\textcolor{red}{XXX Jaebum #1}}
\newcommand{\blue}[1]{{\color{blue}{#1}}} 

\begin{abstract}

Energy efficiency has emerged as a critical challenge in modern base stations (BSs), as the power amplifier (PA) consumes a substantial portion of the total power due to its limited efficiency. We investigate waveform and mode adaptation to enhance the energy efficiency of BSs. We propose Switch-DFT, an adaptive switching framework that selects between cyclic prefix orthogonal frequency division multiplexing (CP-OFDM) and discrete Fourier transform-spread-OFDM (DFT-s-OFDM) waveforms, as well as between single-input multiple-output (SIMO) and multiple-input multiple-output (MIMO) modes. Switch-DFT improves efficiency by reducing PA backoff with DFT-s-OFDM and achieves the target rate at lower power by leveraging higher MIMO throughput. This results in superior energy efficiency over a wide range of the spectral efficiencies compared with static configurations.\\

\end{abstract}


\section{Introduction}

Energy efficiency has become a critical KPI in cellular systems \cite{3gpp:6GWorkshopReport2025}. At base stations (BSs), the power amplifiers (PAs) account for about 40\% of the total power consumption \cite{GreenFutureNetworks2021, GabrielCompletelyNewPower}. Linearization techniques such as digital predistortion (DPD) are widely used to improve PA efficiency~\cite{MorganEtAlGeneralizedMemoryPolynomial2006}. As the radio unit (RU) and distributed unit (DU) become increasingly disaggregated in modern cellular systems, the feedback loop between them suffers from latency and bandwidth constraints, which limit timely DPD adaptation and reduce its effectiveness. The increasingly software-defined DUs employ energy-saving techniques such as single instruction multiple data (SIMD) processing \cite{ParkEtAlAcceleratingVRANORAN2026}, yet operating multiple RUs still consumes significant power due to many PAs. This challenge opens the door for alternative waveform-based solutions. In particular, DFT-s-OFDM, originally adopted in the uplink for its low peak-to-average power ratio (PAPR), offers improved PA efficiency. It can also be an attractive option for the downlink, especially for single-layer transmission. Multi-layer operation involves more complex receiver processing and higher user equipment (UE) cost. This motivates a switching strategy that dynamically selects between DFT-s-OFDM and CP-OFDM, enabling or disabling MIMO functionality to save power consumption.

Prior work has mainly focused on uplink link adaptation and energy-efficient design \cite{ChaeEtAlAdaptiveMIMOTransmission2010, MiaoEnergyEfficientUplinkMultiUser2013}. Early studies examined UEs, where switching between SIMO and MIMO modes reduced UE power consumption~\cite{KimEtAlCrosslayerApproachEnergy2009}. Traditional research compared DFT-s-OFDM and CP-OFDM in the uplink, highlighting PAPR reduction and improved PA efficiency \cite{MyungEtAlSingleCarrierFDMA2006}. More recently, gearbox-PHY tuned PHY parameters for energy savings \cite{GastEtAlHardwareawareEnergyEfficiency2024}, and energy-efficient tri-hybrid MIMO architectures with metasurface antennas have gained traction~\cite{CastellanosEtAlEmbracingReconfigurableAntennas2026}. Inspired by these directions, we develop an energy-efficient approach focusing on RUs within modern BSs.

In this paper, we determine the impact of applying DFT-s-OFDM in the downlink with the PA, focusing on its lower PAPR effects on the signal constellation. We analyze the resulting PA backoff and power-added efficiency (PAE), showing that DFT-s-OFDM enables more energy-efficient SIMO transmission than CP-OFDM. We further propose an energy-efficient RU strategy that dynamically switches between CP-OFDM and DFT-s-OFDM waveforms as well as SIMO and MIMO modes. Simulation results confirm that the proposed method outperforms conventional static configurations in energy efficiency across the spectral efficiency range.

\textit{Notation:} We denote bold lowercase $\ba$ as a vector and uppercase $\bA$ as a matrix. $\ba^T$ and $\bA^H$ denote the transpose and conjugate transpose. We denote $\norm{\bA}$ as the Frobenius norm. We use $\bI_N$ for $N \times N$ identity matrix and $\b0_N$ for a zero vector.

\section{System model with PA analysis}

We first present an analytical DFT-s-OFDM system model that incorporates PA nonlinearity. To highlight the nonlinear behavior of the PA, we assume a SISO scenario and a simplified operating regime that isolates the impact of PA distortion. Although DFT-s-OFDM offers limited resource allocation flexibility in multi-user downlink scenarios compared to CP-OFDM, it can be employed as a fallback and optional transmission mode under specific conditions, such as single-user scheduling. The analysis in this section focuses on such scenarios. The generalization to a MIMO system is also straightforward by concatenating the per-antenna SISO channel matrices into a block matrix.

We begin by defining the transmitted signal of a single-user DFT-s-OFDM system. The data symbol vector $\bd \in \bbC^{M \times 1}$ contains QAM-modulated symbols with unit average symbol power. Let $M$ denote the number of resource elements (REs) allocated to the user and $N$ the FFT size. The $M \times M$ DFT matrix $\bF_M$ is defined as
\begin{equation}\label{eq:tx_vector}
    \left[ \bF_M \right]_{\ell,m} = \frac{1}{\sqrt{M}} 
    \exp \left( -j \frac{2\pi \ell m}{M} \right), \quad \forall \ell,m \in \{0,1,\ldots,M-1\}. 
\end{equation}
The ${N \times M}$ resource mapping matrix ${\bT} \in \{0,1\}$ allocates the DFT-spread symbols ${\bF}_M {\bd}$ to $N$ subcarriers using a split-localized mapping to lower the PAPR. For CP-OFDM, we set $\bF_{M}$ to $\bI_{M}$ and directly map the symbols to the frequency grid. The time-domain transmit signal vector $\bx \in \bbC^{N \times 1}$ can be written as
\begin{equation} \label{eq:time_dft}
     \bx = \bF_{N}^{H} \bT \bF_M \bd.
\end{equation}
Assuming CP insertion and removal, we omit the explicit CP operators for notational simplicity here. The received time-domain signal vector $\by \in \bbC^{N \times 1}$ and the received frequency-domain signal vector ${\bY} \in \bbC^{N \times 1}$ are
\begin{align}
    {\by} &= {\bH} {\bF}_N^H {\bT} {\bF}_M {\bd} + {\bv}, \label{eq:rx_time_vector}\\
    \bY &= \bF_N \by = \bF_N {\bH} {\bF}_N^H {\bT} {\bF}_M {\bd} + \bF_N {\bv}. \label{eq:rx_freq_vector}
\end{align}
The channel matrix ${\bH} \in \bbC^{N \times N}$ is circulant, and its $n$-th column is a circularly shifted version of the zero-padded impulse response, given by
\begin{equation}
    \begin{aligned}\label{eq:ch_matrix}
      {\left[ {\bH} \right]_{\left( {:, } n \right)}} = {\mathrm{circshift}}\left\{ {{{\left[ {{h_0}\, \cdots \,{h_{{L} - 1}}\,\,{{\bf{0}}_{N - {L}}}} \right]}^T},\,n} \right\}, \\ \forall n \in \{0,1,\ldots,N-1\}, 
    \end{aligned}
\end{equation}
where $\bv \sim \cN_{\bbC}(\b0_N, \sigma_{\sfN}^{2}\bI_N)$ is the AWGN vector with noise variance $\sigma_{\sfN}^{2}$.

We analytically model the PA nonlinearity using a memoryless discrete time formulation based on the modified Rapp model, which captures both amplitude-to-amplitude modulation (AM/AM) and the amplitude-to-phase modulation (AM/PM) \cite{RappPA1991, DudakKahyaogluDescriptiveStudyAMAM2012a}. To maintain analytical tractability, we omit oversampling and model the PA using discrete time DFT-s-OFDM samples at the Nyquist rate over each OFDM symbol interval. This allow us to exclude the impact of PA nonlinearity on the adjacent channel leakage ratio (ACLR) and to focus on in-band performance metrics such as error vector magnitude (EVM).

Let $x[n]$ denote the $n$-th Nyquist-rate sample of the baseband input signal $\bx$ to the PA within one OFDM symbol duration. The PA is modeled by a nonlinear complex gain function $q(\cdot): \bbC \rightarrow \bbC$ that depends on the instantaneous amplitude of its input, defined as 
\begin{equation}
   q(x[n]) = g(|x[n]|)\, e^{j\,\theta(|x[n]|)}.
\end{equation}
The functions $g(\cdot)$ and $\theta(\cdot)$, following the modified Rapp model \cite{DudakKahyaogluDescriptiveStudyAMAM2012a}, represent the AM/AM and AM/PM, where
\begin{equation}
\begin{aligned}
    g(|x[n]|) &= |x[n]|\!\left(1 + \!\left(\tfrac{|x[n]|}{A_{\mathrm{sat}}}\right)^{2s}\right)^{-1/(2s)}, \\
\theta(|x[n]|) &= \alpha\,|x[n]|^{q_1}\left(1 + (\tfrac{|x[n]|}{\beta})^{q_2}\right)^{-1}, 
\end{aligned}
\end{equation}
$A_{\text{sat}}$ is the saturation amplitude, $s$ is the smoothness factor of the Rapp model, and $\alpha$, $\beta$, $q_1$, and $q_2$ are empirical fitting parameters for modeling the measured AM/PM characteristics of practical PAs.

\begin{figure}[tbp]
\centering
\includegraphics[draft=false,width=0.49\columnwidth] {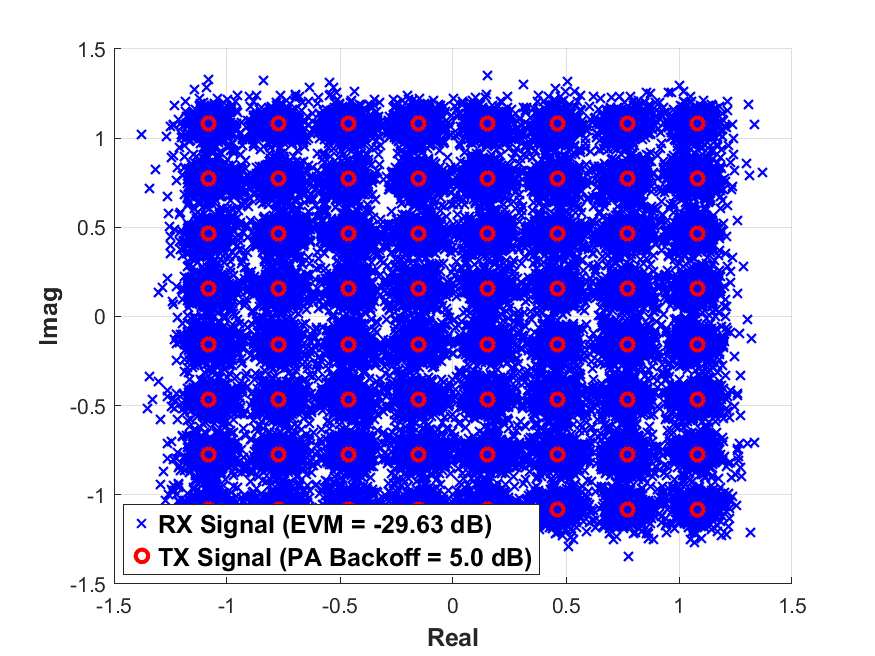}
\includegraphics[draft=false,width=0.49\columnwidth] {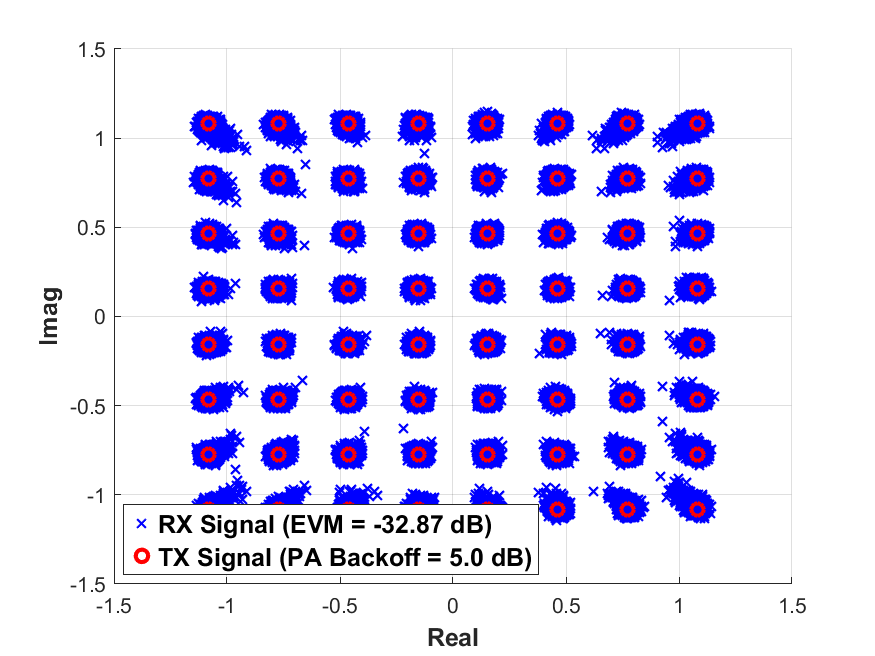}
\caption{64QAM constellations of CP-OFDM (left) and DFT-s-OFDM (right) with PA nonlinearity (TDL-C channel, PA backoff $=5$~dB). Compared with CP-OFDM, the lower PAPR yields more compact received signal constellation clusters and about 3~dB EVM improvement.}
\label{const_cp_dft}
\end{figure}

\begin{figure}[t]
    \centering
    \includegraphics[width=0.8\linewidth]{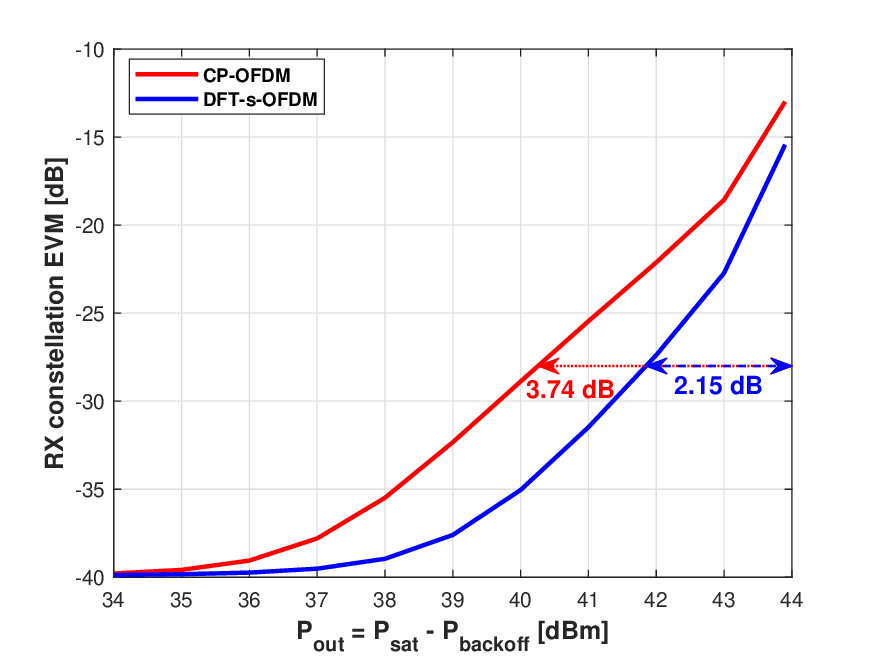}
    \caption{EVM versus PA backoff for CP-OFDM and DFT-s-OFDM. 
    DFT-s-OFDM requires less backoff to satisfy a given EVM constraint (-28 dB), enabling higher transmit power and improved PA efficiency.}
    \label{EVM_backoff}
\end{figure}

Let $\bQ(\bx) ={\mathrm{diag}} \{ q(x[0]), q(x[1]), \ldots, q(x[{N-1}]) \}$ denote a diagonal operator whose entries multiply the corresponding input samples. The time-domain PA output vector $\bx^{\PA}$ is then written as $\bx^{\PA} = \bQ(\bx) \bx$. By substituting this into \eqref{eq:time_dft} and \eqref{eq:rx_freq_vector}, the received frequency-domain  signal vector including PA nonlinearity becomes 
\begin{equation}
\begin{aligned}\label{eq:rx_freq_pa}
{\bY} = {\bF_N}{\bH}{\bQ(\bx)}{\bx} + {\bF_N}{\bv} = {{\bF_N}{\bH}{\bQ(\bx)}{\bF}_N^H}{\bT}{\bF}_M{\bd} + {\bF_N}{\bv}.
\end{aligned} 
\end{equation}
From \eqref{eq:rx_freq_pa}, we define the effective channel as $\bH_{\mathrm{eff}} = {{\bF_N}{\bH}{\bQ(\bx)}{\bF}_N^H}$. Then, the equalized data symbol vector $\hat{\bd} \in \bbC^{M \times 1}$ obtained using a linear MMSE equalizer is
\begin{equation}\label{eq:rx_dhat}
\begin{aligned}
    \hat{\bf{d}}  = &{\bF}_M^H \bH_{\mathrm{eff}}^{H} (\bH_{\mathrm{eff}}\bH_{\mathrm{eff}}^{H}+\sigma_{\sfN}^{2}\bI)^{-1} \bH_{\mathrm{eff}} {\bT} {\bF}_M {\bd} \\
       &+ {\bF}_M^H \bH_{\mathrm{eff}}^{H} (\bH_{\mathrm{eff}}\bH_{\mathrm{eff}}^{H}+\sigma_{\sfN}^{2}\bI)^{-1} {\bF}_N {\bv}.
\end{aligned}
\end{equation}
To provide intuition on the impact of PA nonlinearity, \figref{const_cp_dft} illustrates the constellations of CP-OFDM and DFT-s-OFDM under identical PA conditions. The plots compare the transmitted symbols $\bd$ with the equalized symbols $\hat \bd$, indicating the EVM. Since DFT-s-OFDM has a lower PAPR, it achieves about 3 dB lower EVM than CP-OFDM at the same backoff. \figref{EVM_backoff} shows that, at a given PA saturation power $P_\sat$, meeting the EVM requirement requires less backoff $P_{\text{backoff}}$ with DFT-s-OFDM. This enables a higher transmit power $P_{\text{out}}=P_{\sat}-P_{\text{backoff}}$ and thus improved PA efficiency. CP-OFDM exhibits higher sensitivity to nonlinear distortion due to its larger PAPR, whereas DFT-s-OFDM maintains more robust constellation points with lower EVM under the same PA characteristics. 

We determine the PA backoff and PAE that ensure approximately linear transmitter operation. These parameters are incorporated into the RU power consumption model analyzed in the next section. Under this condition, applying a Bussgang decomposition  \cite{BussgangCrosscorrelationFunctionsAmplitudedistorted1952a}, the nonlinear PA output in \eqref{eq:rx_freq_pa} can be expressed as ${\bQ(\bx)}{\bx} = a \bx + \bd$, where $\bd$ denotes the distortion uncorrelated with the input $\bx$. With a sufficiently large backoff, the distortion becomes negligible ($\bd \approx \b0$), yielding an equivalent linear channel $\by \approx a \bH \bx + \bv$. Therefore, the mutual information expression in the next section approximates the achievable spectral efficiency.

\section{Energy-efficient and power consumption analysis of radio unit}

We analyze the RU power consumption in SIMO and MIMO settings, and quantify its tradeoff with spectral efficiency. We also examine how the tradeoff changes when DFT-s-OFDM is applied in the downlink SIMO transmission, showing that lower PAPR improves PA efficiency and shifts the crossover point between SIMO and MIMO. Although time-division duplexing (TDD) systems include both downlink and uplink durations, we focus on the downlink transmission in this paper.

\subsection{Transmit power and spectral efficiency in SIMO and MIMO}

We first analyze the relationship between transmit power and spectral efficiency for SIMO and MIMO systems. We assume no channel state information at the transmitter (CSIT), and therefore adopt equal power allocation without precoding in this paper. With CSIT, water-filling or antenna muting can further improve efficiency, which we leave for future work. 

Let $\lambda_j(\bH^H \bH)$ denote the $j$-th eigenvalue and the number of data streams be equal to the number of transmit antenna, i.e., $N_\rms = N_\rmt = \min(N_\rmt, N_\rmr)$. The achievable spectral efficiency of MIMO is expressed as \cite{HeathLozanoFoundationsMIMOCommunications2018}
\begin{equation}\label{eq:mimo_capacity}
    C = \sum_{j=0}^{N_\rmt-1} \log_2 \left( 1 + \frac{\SNR}{N_\rmt} \, \lambda_j \!\left( \bH^H\bH \right) \right).
\end{equation}
Let $G$ denote the large-scale channel gain, $E_\rms$ the transmit energy per resource element, $B$ the bandwidth, $P_{\TX}$ the PA's output power that is dissipated into the air, and $N_0$ the noise power spectral density. The received signal-to-noise ratio (SNR) is then written as $\SNR = G E_\rms/N_0 = G P_\TX/(N_0B)$.

For a $2 \times 2$ MIMO system, \eqref{eq:mimo_capacity} can be expanded as 
\begin{equation} \label{eq:mimo_expand}
 \begin{aligned}
   C  &= \log_2\!\left(\prod_{j=0}^{1} \left(1 + \frac{G\lambda_j}{2N_0B}\,P_{\TX}\right)\right) \\
      &= \log_2\!\left(  \lambda_0 \lambda_1 \left(\frac{G}{2N_0B}\right)^2 P_{\TX}^2 + (\lambda_0+\lambda_1) \left(\frac{G}{2N_0B}\right) P_{\TX} + 1  \right).  
\end{aligned}   
\end{equation}
Rearranging \eqref{eq:mimo_expand} yields a quadratic equation for $P_{\TX}$ given by
\begin{equation} \label{eq:mimo_quad}
   \lambda_0 \lambda_1 \left(\frac{G}{2N_0B}\right)^2 P_{\TX}^2 + (\lambda_0+\lambda_1) \left(\frac{G}{2N_0B}\right) P_{\TX} + (1-2^C) = 0.
\end{equation}
Among the two solutions in \eqref{eq:mimo_quad}, the feasible MIMO transmit power is given by the positive root as
\begin{equation}\label{eq:mimo_ptx}
    P_{\TX}^{\text{MIMO}} 
    = \frac{2N_0B}{G\lambda_0\lambda_1}\left( {
    \sqrt{\left(\tfrac{\lambda_0+\lambda_1}{2}\right)^2 
    + \lambda_0 \lambda_1 \big(2^C-1\big) } - \frac{\lambda_0+\lambda_1}{2}} \right).
\end{equation}

For a single transmit antenna ($N_\rmt=1$ in~\eqref{eq:mimo_capacity}), we can define the effective SIMO gain as $\lambda_{\mathrm{eff}}=\max_{1\le j\le N_\rmt}\|\bh_j\|^2$. In this case, the UE measures $\|\bh_j\|^2$ for all $N_{\rmt}$ transmit antennas and feeds back the index of the best antenna, which requires $\log_2 N_{\rmt}$ bits of feedback. The decision is made at the BS, leveraging limited feedback from the UE. The required SIMO transmit power for spectral efficiency $C$ is
\begin{equation}\label{eq:simo_ptx}
  P_{\TX}^{\text{SIMO}} 
  =\frac{N_0 B}{G\,\lambda_{\mathrm{eff}}}\,\big(2^{C}-1\big).
\end{equation}

\subsection{RU power consumption and crossover point analysis with DFT-s-OFDM}

Based on the transmit power expressions, we model the total RU power consumption as the sum of the PA DC power and the circuit power. Let $M_{\mathrm{act}}$ denote the number of active PA chains, $P_{\PA,i}$ the DC power drawn by the $i$-th PA, and $P_{\mathrm{circ}}$ the static circuit power. The total RU power $P_\RU$ is written as 
\begin{equation} \label{eq:RU_power}
    P_{\mathrm{RU}} = \sum_{i=1}^{M_{\mathrm{act}}} P_{\PA,i} + P_{\mathrm{circ}}.
\end{equation}

The PA DC power follows a drain-efficiency model. Given the $i$-th PA’s TX output power $P_{\TX,i}$, saturation output power $P_{\sat,i}$, output power backoff $b$ (dB), and drain efficiency $\eta_i(b)$, the DC power is $P_{\PA,i}=P_{\TX,i}/\eta_i(b)$, and under backoff the TX output satisfies $P_{\TX,i}=P_{\mathrm{sat},i}10^{-b/10}$.

The circuit power $P_{\mathrm{circ}}$ accounts for the RF front-end chain, mainly from several hardware components. Let $P_{\LO}$ denote the local oscillator (LO) power, $P_{\filt}$ the filter power, $P_{\mix}$ the mixer power, $P_{\dac}$ the digital-to-analog converter (DAC) power, and $P_{\PA,\idle}$ the idle biasing power of the PA. With these definitions, the circuit power is expressed as $P_{\mathrm{circ}} = P_{\LO} + M_{\mathrm{act}} \, (P_{\filt} + P_{\mix} + P_{\dac} + P_{\PA,\idle})$.

\begin{figure}[t]
    \centering
    \includegraphics[width=0.8\linewidth]{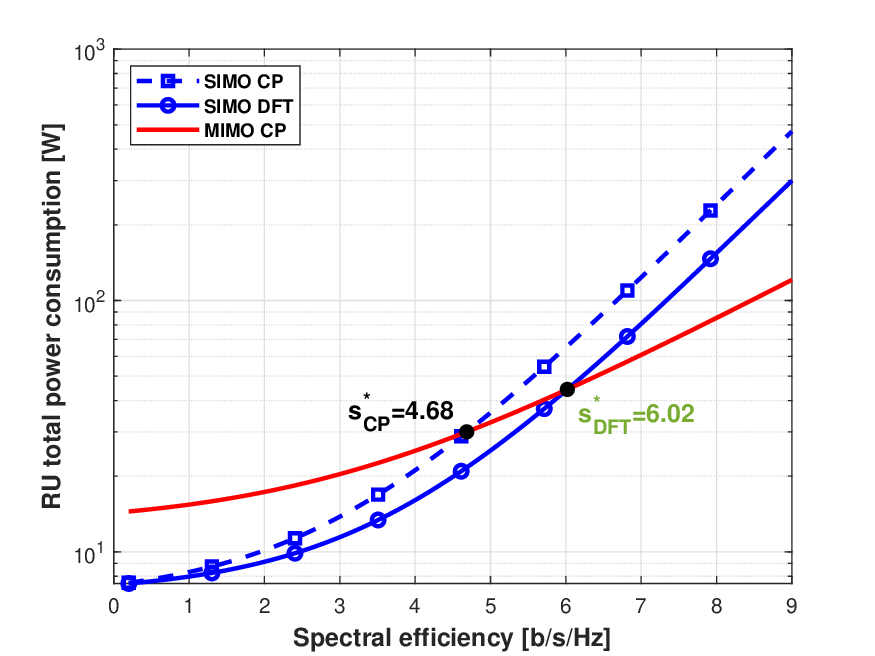}
    \caption{Total RU power consumption versus spectral efficiency for SIMO (CP-OFDM or DFT-s-OFDM) and MIMO (CP-OFDM). Using DFT-s-OFDM for SIMO transmission reduces power consumption and extends the spectral efficiency range where SIMO remains preferable.}
    \label{crossover}
\end{figure}

By substituting \eqref{eq:mimo_ptx} and \eqref{eq:simo_ptx} into \eqref{eq:RU_power} and equating the two expressions, we can identify the \emph{\textbf{crossover point}} $(s^*,p^*)$, defined as the spectral efficiency $s^*$ at which the RU power consumption $p^*$ of SIMO and MIMO becomes equal. Below this point, SIMO achieves the target rate with lower power, whereas above it MIMO becomes more power-efficient due to its multiplexing gain. This crossover-based analysis, originally studied for UE power consumption \cite{KimEtAlCrosslayerApproachEnergy2009}, can be extended to the BS by including different waveforms and PA efficiency effects. When DFT-s-OFDM is applied, the reduced PAPR improves PA efficiency and shifts the crossover point towards higher spectral efficiency. As shown in \figref{crossover}, SIMO with DFT-s-OFDM remains preferable over a wider operating range.

\subsection{Energy-efficiency optimization problem}

Let $R_{\mathrm{eff}}$ denote the effective throughput (data rates). The RU’s energy efficiency (EE) is defined as \cite{YazdanEtAlEnergyefficientMassiveMIMO2017}
\begin{equation}
\label{eq:ee-def}
\mathrm{EE} = \frac{R_{\mathrm{eff}}}{P_\RU}.
\end{equation}
The effective throughput is given by the product of the transmission bandwidth $B$ and the achievable spectral efficiency $S$, scaled by an overhead factor $\alpha_{\mathrm{OH}}\in(0,1]$ that accounts for cyclic prefix, pilots, and control signaling. Thus, the effective throughput is expressed as $R_{\mathrm{eff}} = \alpha_{\mathrm{OH}} BS$.

Based on these definitions, we formulate the RU EE maximization problem. Let $m$ denote the transmission mode, and $f(P_{\TX,i}^{(m)})$ the corresponding EE function as 
\begin{equation} \label{eq:ee_P_tx}
    f(P_{\TX,i}^{(m)}) = \frac{ \alpha_{\mathrm{OH}} B \sum_{i=1}^{M_{\mathrm{act}}(m)}\log_ 2\left(1+\frac{G\,\lambda_{i}^{(m)}} {N_0 B} P_{\TX,i}^{(m)}\right)}
    { \sum_{i=1}^{M_{\mathrm{act}}(m)} \frac{P_{\TX,i}^{(m)}}{\eta_m(b)} \;+\; P_{\mathrm{circ}}(m)}.
\end{equation}
The optimization problem of \eqref{eq:ee_P_tx} is then written as
\begin{equation} \label{eq:objective}
    \max_{m, b , P_{\TX,i}^{(m)}} f(P_{\TX,i}^{(m)}) ,
\end{equation}
\vspace{-6mm}
\begin{subequations}\label{eq:consts}
\begin{align}
\text{s.t.} \quad
m \in \{\mathsf{Full}\!-\!\mathsf{MI}&\mathsf{MO}, 
\mathsf{Switch\!-\!CP}, \mathsf{Switch\!-\!DFT}\}, \label{eq:opt-mode} \\
& b \ge b_{\min}^{\mathrm{EVM}}(m,M_{\mathrm{QAM}}), \label{eq:opt-evm-bo} \\ 
& P_{\TX,i}^{(m)} \ge \frac{P_{\mathrm{req}}(m)}{M_{\mathrm{act}}(m)}, \label{eq:opt-minptx} \\
& P_{\TX,i}^{(m)} \le  P_{\sat,i}  10^{-b/10}. \label{eq:opt-ptxcap}
\end{align}
\end{subequations}

\noindent The transmission mode $m$ in \eqref{eq:opt-mode} is defined as follows. \textbf{Full-MIMO} uses MIMO with CP-OFDM only. \textbf{Switch-CP} switches between SIMO and MIMO, both with CP-OFDM. \textbf{Switch-DFT} uses SIMO with DFT-s-OFDM and MIMO with CP-OFDM. The constraint in \eqref{eq:opt-evm-bo} enforces the minimum backoff $b_{\min}^{\EVM}$ required to satisfy the modulation accuracy for a given QAM order, defined as
\begin{equation}
b_{\min}^{\EVM}(m,M_{\mathrm{QAM}}) = 
\inf\big\{b:\ \EVM_m(b)\le \EVM_{\mathrm{req}}(M_{\mathrm{QAM}})\big\}.
\nonumber
\end{equation}
Constraint \eqref{eq:opt-minptx} ensures that the per-antenna transmit power is no smaller than the minimum required to satisfy the target spectral efficiency. \eqref{eq:opt-ptxcap} limits the transmit power according to the PA saturation under the applied backoff. These constraints ensure that the optimization problem reflects both system-level requirements and device-level limitations.

\section{Single-ratio fractional programming optimization}

We consider a single-ratio fractional programming (FP) optimization problem as a function of the transmit power variable $p = P_{\TX,i}$ under equal power allocation. We then simplify \eqref{eq:ee_P_tx} as $f(p) = A(p)/B(p)$, where the numerator is given by $A(p) = \alpha_{\mathrm{OH}} B\log(1+cp)$, and the denominator by $B(p) = p/\eta+P_{\mathrm{circ}}$. For fixed $b$ and $m$, we reformulate the RU objective function in \eqref{eq:objective} as  
\begin{equation}
\label{eq:fp-original}
\max_{P_{\min} \le p \le P_{\max}}\; f(p),
\end{equation}
where $c > 0$ denotes the effective channel gain normalized by the noise power. The numerator $A(p)$ is concave and increasing, while the denominator $B(p)$ is affine and strictly positive for $P_{\mathrm{circ}}>0$. Unless stated otherwise, $\log$ denotes the natural logarithm. Replacing $\log$ with $\log_2$ scales $A(p)$ by \(1/\ln 2\) and does not affect the optimizer. For the SIMO case, the achievable spectral efficiency is $S_{\mathrm{SIMO}}(p)=\log_2(1+cp)$, which is concave. For the $2\times2$ MIMO case, $S_{\mathrm{MIMO}}(p)=\log_2\!\big((1+c_0 p)(1+c_1p)\big)=\log_2(1+c_0p)+\log_2(1+c_1 p)$, is a sum of concave functions. These properties ensure a single-ratio concave-convex FP structure for global optimization via the quadratic transform.

\subsection{Equivalent quadratic transform}
Applying the quadratic transform~\cite{ShenYuFractionalProgrammingCommunication2018}, the ratio $f(p)$ can be expressed by the bivariate objective as
\begin{equation} \label{eq:quad_transform}
g(p,y) =  -B(p) y^2 + 2y\sqrt{A(p)}.
\end{equation}
For a fixed $p$, the optimal $y$ is obtained in closed form as 
\begin{equation} \label{eq:y-update}
y^\star(p) = \frac{\sqrt{A(p)}}{B(p)} \;\ge\; 0,
\end{equation}
and substituting this value back into \eqref{eq:quad_transform} recovers the original,
\begin{equation}
    \max_{y\in\bbR} g(p,y) = \frac{A(p)}{B(p)}.
\end{equation}
Therefore, maximizing  $f(p)$ over $p$ is equivalent to maximizing $g(p,y)$ jointly over $(p,y)$.

\begin{algorithm}[!b]
\caption{Quadratic transform for maximizing RU EE}
\label{alg:fpsolution}
\begin{algorithmic} 
\STATE \textbf{Step 0:} Initialize $k=0$ and $p^{(0)} \in [P_{\min}, P_{\max}]$.
\STATE \textbf{Repeat}
\STATE \quad \textbf{Step 1:} Update $y^{(k)}=\sqrt{A(p^{(k)})}/{B(p^{(k)})}$.
\STATE \quad \textbf{Step 2:} Find $\hat p$ satisfying $\tfrac{\partial}{\partial p} g(\hat p, y^{(k)}) = 0$.
\STATE \quad \textbf{Step 3:} Update $p^{(k+1)}=\hat p$.
\STATE \quad \textbf{Step 4:} Increase $k$ by one.
\STATE \textbf{Until} $\lvert p^{(k)} - p^{(k-1)} \rvert \le \varepsilon$.
\STATE \textbf{Output:} $p^\star = p^{(k)}$, $y^\star = y^{(k)}$, and $f(p^\star) = A(p^\star)/B(p^\star)$.
\end{algorithmic}
\end{algorithm}

\subsection{Subproblem in $p$}
The quadratic transform enables alternating optimization over $p$ and $y$. The $y$-update admits a closed-form solution, whereas the $p$-update requires solving a concave maximization problem. For fixed $y \ge 0$, the objective is $g(p,y) = 2y\sqrt{A(p)} - y^2 B(p)$, which is concave since $\sqrt{A(p)}$ is concave and $-y^2 B(p)$ is affine. Therefore, the global maximizer can be obtained by solving the first-order optimality condition.

As an example, we consider the SIMO case with $A(p)=\alpha_{\mathrm{OH}}B\log(1+cp)$ and $B(p)=p/\eta+P_{\mathrm{circ}}$. The global maximizer is obtained by solving 
\begin{equation} \label{eq:FOC}
\frac{\partial}{\partial p}g(p,y) = y\frac{A'(p)}{\sqrt{A(p)}} - y^2 B'(p) = 0,
\end{equation}
where $A'(p)=\alpha_{\mathrm{OH}} B\,\dfrac{c}{1+cp}$ and $B'(p)=1/\eta$. 
Substituting these expressions reduces \eqref{eq:FOC} to 
\begin{equation} \label{eq:foc_root}
\frac{c\sqrt{\alpha_{\mathrm{OH}} B}}{(1+cp)\sqrt{\log(1+cp)}} = \frac{y}{\eta}, \qquad y>0.
\end{equation}
This has a unique solution $\hat p(y)$ on $[P_{\min},P_{\max}]$ since the left-hand side is strictly decreasing, which can be verified by showing its derivative is negative. The $p$-update is obtained by solving this equation for $\hat p(y)$, and combined with the $y$–update to yield the full iterative procedure.

\subsection{Iterative algorithm and global convergence}
We solve \eqref{eq:fp-original} iteratively via alternating optimization with the quadratic transform, updating $y$ in the closed form and $p$ by one-dimensional search. The algorithm, initialized at $p^{(0)}=\tfrac{P_{\min}+P_{\max}}{2}$, is summarized in \algref{alg:fpsolution}. In the single–ratio case, the quadratic transform introduces no spurious stationary points, so the stationary points of $g(p,y)$ and $f(p)$ coincide. Since $g(\cdot,y)$ is concave in $p$ for all $y\ge 0$ and the $y$–update is globally optimal for fixed $p$, the iterations generate a monotonically nondecreasing sequence $\{f(p^{(k)})\}$ that converges to the global maximizer $(p^\star,y^\star)$ with objective $f(p^\star) = \log(1+cp^\star)/(p^\star/\eta + P_{\mathrm{circ}})$.
This result highlights that the optimal point $p^\star$ balances the transmission-rate increase in the numerator with the PA and circuit power in the denominator, thereby maximizing the energy efficiency.

\begin{table}[!b]
    \caption{Assumptions and parameters for EE optimization.\label{tbl:EE_opt_params}}
    \begin{center}
        {\renewcommand{\arraystretch}{1.3}%
            \begin{tabular} {cc|cc}
            \hline\hline
            & Parameter & Value &\\
            \hline
            & Waveform & CP-OFDM / DFT-s-OFDM &\\
            & Carrier frequency & 3.5 Hz &\\
            & Subcarrier spacing & 15 kHz &\\
            & Resource block & 100 &\\
            & DFT size $M$ & 1200 &\\
            & FFT size $N$ & 2048 &\\
            & Bandwidth $B$ & 20 MHz &\\
            & Modulation & 64QAM &\\            
            & Channel model & TDL-C, 100km/h &\\
            & Transmission overhead & 0.9 &\\
            & Min. EVM requirement & -31 dB &\\
            \hline\hline
            \end{tabular}
        }
    \end{center}
    \vspace{-2mm}
\end{table}

\section{Energy efficiency and power consumption results}

We now present simulation results to evaluate the RU’s energy efficiency (EE) improvement achieved by the adaptive \textbf{Switch-DFT} mode. For comparison, we use \textit{Full SIMO} as the baseline. In the switching approach, the EE of SIMO and MIMO is first optimized separately, and the system selects the mode with the higher fractional-programming (FP)-based optimal EE. Switching is implemented by simply comparing the two results, and the decision typically changes near the crossover region.

For the numerical evaluation, we adopt practical RF parameters from the Qorvo QPA3505 PA, designed for 3.5 GHz 5G massive MIMO BSs, with $P_\sat = 44$ dBm and $P_\PA = 9.15$ W \cite{pa_qpa3503}. The static power consumption of the RF circuit components is set as follows: $P_{\dac} = 3.5$ W, $P_{\LO} = 0.5$ W, $P_{\mix} = 0.38$ W, $P_{\filt} = 0.02$ W, and $P_{\PA,\idle} = 3.5$ W. For the modified Rapp model, the parameters are set to $s = 3$, $\alpha = 190\pi/180$, $\beta = 0.1$, $q_1 =3.8$ and $q_2 = 2.5$.

First, the link-level simulation parameters are summarized in \tabref{tbl:EE_opt_params}, based on the 3GPP specifications \cite{dahlman20205g}. We consider a $2 \times 2$ MIMO configuration that can fall back to SIMO via single-antenna selection. The transmission overhead, the minimum EVM requirements for QAM modulations (with margin), and the TDL-C channel model are all based on the 3GPP BS and UE standards \cite{dahlman20205g}.

\begin{figure}[t]
    \centering
    \includegraphics[width=0.9\linewidth]{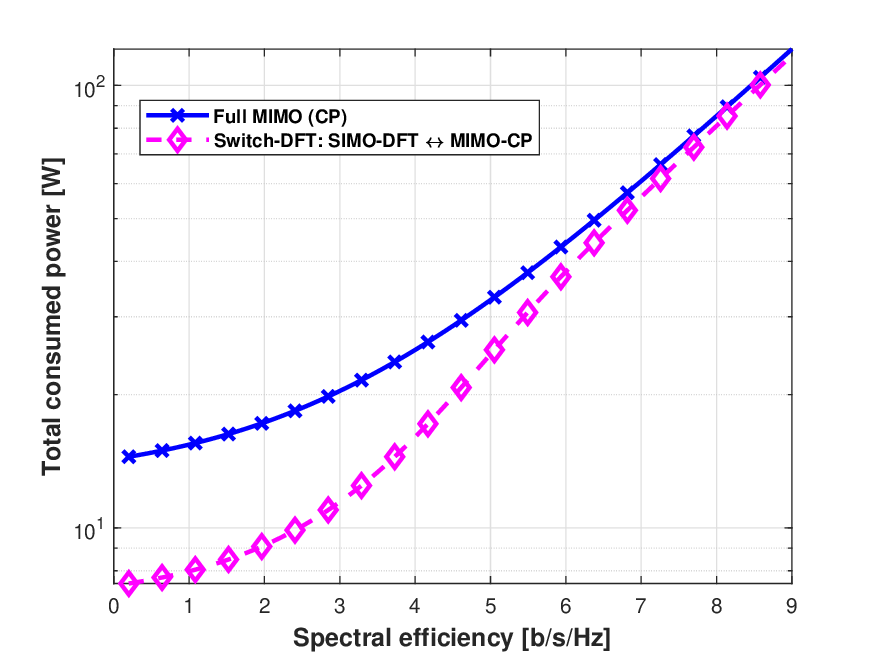}
    \caption{Total RU power consumption versus spectral efficiency for different transmission modes. The proposed Switch-DFT strategy achieves the lowest power consumption across the entire range and outperforms the baseline.}
    \label{opt_TCP_SE}
\end{figure}

\begin{figure}[t]
    \centering
    \includegraphics[width=0.9\linewidth]{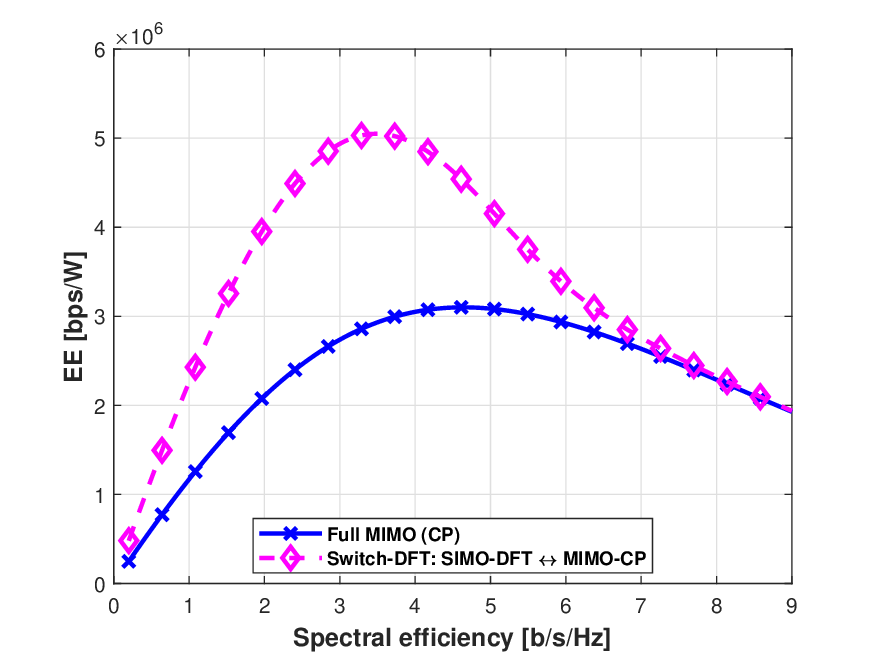}
    \caption{Energy efficiency (EE) versus spectral efficiency for SIMO and MIMO with CP-OFDM and DFT-s-OFDM. The proposed Switch-DFT achieves higher EE across both low and high spectral efficiency regions, outperforming the baseline.}
    \label{opt_EE_SE}
\end{figure}

In \figref{opt_TCP_SE}, the proposed Switch-DFT mode shows the lowest power consumption across all spectral efficiency values. This indicates that it provides a more optimal operating point compared to the baseline Full MIMO (CP) mode. \figref{opt_EE_SE} illustrates the energy efficiency results obtained via FP optimization. Again, the Switch-DFT mode achieves the highest EE compared to the Switch-CP mode and Full MIMO mode. It is worth noting that the optimal spectral efficiency point of Switch-DFT appears to the left of that of Full MIMO. This small shift occurs because Switch-DFT benefits from reduced PA backoff and lower circuit power consumption, which boost EE in the low spectral efficiency region. As a result, the EE curve of Switch-DFT reaches its maximum earlier. In other words, Switch-DFT achieves peak efficiency at lower rates while still maintaining higher EE than Full MIMO mode.

\section{Conclusion}
In this paper, we presented an energy-efficient framework for RUs that integrates adaptive waveform selection and mode switching. By analytically modeling PA nonlinearity and circuit power consumption, we demonstrated that the proposed Switch-DFT strategy significantly improves energy efficiency compared with static CP-OFDM/MIMO configurations. The results highlight the benefit of dynamically exploiting both the low-PAPR characteristics of SIMO DFT-s-OFDM and the capacity gains of MIMO. Although the current framework employs a memoryless Rapp PA model and a 2×2 MIMO configuration, the findings confirm that adaptive switching is a promising path for next-generation RANs. Future work will extend this study to memory PA models, large-scale MIMO, RU sleep modes, traffic-aware scheduling, and reinforcement learning-based real-time control to further enhance the practicality of energy-efficient RAN operation.

\section{Acknowledgment}
The work of Robert W. Heath Jr. was supported by the National Science Foundation under grant nos. NSF-ECCS-2435261, NSF-CCF-2435254, NSF-ECCS-2414678 and in part by the Army Research Office under Grant W911NF2410107. The work of C.-B. Chae was supported in part by Korean government (RS-2024-00428780 and RS-2026-25489110).

\bibliographystyle{IEEEtran}
\bibliography{JP_EE_MIMO_refs}

\end{document}